\documentclass[twocolumn,prl,aps,floatfix]{revtex4}
\usepackage{amsmath,amssymb}
\usepackage{graphicx,graphics}
\usepackage{fancyhdr}

\begin{document}

\title{Mesoscopic Interplay of Superconductivity and \\ Ferromagnetism in Ultra-small Metallic Grains}
\author{S. Schmidt\email{schmidts@phys.ethz.ch}}
\affiliation{Institute for Theoretical Physics, ETH Zurich, 8093 Zurich, Switzerland }
\author{Y. Alhassid\email{yoram.alhassid@yale.edu}}
\affiliation{Center for Theoretical Physics, Sloane Physics
 Laboratory, Yale University,  New Haven, Connecticut 06520, USA}

\begin{abstract}
We review the effects of electron-electron interactions on the ground-state spin and the transport properties of ultra-small chaotic metallic grains. Our studies are based on an effective Hamiltonian that combines a superconducting BCS-like term and a ferromagnetic Stoner-like term. Such terms originate in pairing and spin exchange correlations, respectively. This description is valid in the limit of a large dimensionless Thouless conductance. We present the ground-state phase diagram in the fluctuation-dominated regime where the single-particle mean level spacing is comparable to the bulk BCS pairing gap. This phase diagram contains a regime in which pairing and spin exchange correlations coexist in the ground-state wave function. We discuss the calculation of the tunneling conductance for an almost-isolated grain in the Coulomb-blockade regime, and present measurable signatures of the competition between superconductivity and ferromagnetism in the mesoscopic fluctuations of the conductance.
\end{abstract}

\maketitle

\section{Introduction}

Superconductivity and ferromagnetism compete with each other. Pairing correlations lead to Cooper pairs of electrons with opposite spins and thus tend to minimize the total spin of the grain, while ferromagnetic correlations tend to maximize the total spin.

Nevertheless, it is well known that superconducting and ferromagnetic order can be present simultaneously in bulk systems when ferromagnetism is caused by localized paramagnetic impurities~\cite{AG60,Cl62,Ch62,FF64,LO64}. Recently, it was observed that both states of matter can coexist in high-Tc superconductors~\cite{Ta99,Be99} and in heavy fermion systems~\cite{Sa00,Pf01,Ao01} even when the electrons that are responsible for superconductivity and ferromagnetism are the same. This surprising observation led to the search of new theoretical models that can describe this coexistence.

In ultra-small metallic grains, in which the bulk pairing gap $\Delta$ is comparable to the single-particle mean level spacing $\delta$, a coexistence regime of superconductivity and ferromagnetism was predicted~\cite{falci,ying,schmidt}. The ground state of the grain is described by a state where a few single-particle levels around the Fermi energy are singly occupied while all other electron are paired. This coexistence regime is characterized by spin jumps and its size can be tuned by an external Zeeman field.

However, it is difficult to measure the ground-state spin of a grain, and a more directly measurable quantity is the tunneling conductance through the grain~\cite{delft01}. In addition, one has to take into account the mesoscopic fluctuations that are typical for chaotic grains~\cite{al00}.  Effects of exchange correlations on the conductance statistics in quantum dots, in which pairing correlations are absent, were studied in Ref.~\cite{rupp}. In Ref.~\cite{schmidt2} we identified signatures of the coexistence of pairing and exchange correlations in the mesoscopic fluctuations of the conductance through a metallic grain that is weakly coupled to leads.

The fabrication and control of nano-size metallic devices is a challenging task. The first conductance measurements in ultra-small metallic grains were carried out in the mid-nineties~\cite{RBT95,RBT96,RBT97}. The grains were produced by breaking nanowires and their size was difficult to control. Coulomb blockade, discrete levels and pairing effects were observed in a single
grain by measuring the tunneling conductance~\cite{delft01}. During the last decade numerous technological advances led to an increase in control and tunability of ultra-small metallic grains.  Break junction techniques~\cite{park} and electromigration~\cite{bolotin1} were used for gating and establishing precise contact between leads and grain. A particularly important recent development has been the use of monolayers of organic molecules as tunnel barriers, enabling control of the size and shape of the grain \cite{kuemmeth}.  New materials have been tested as well. Cobalt nanoparticles were used to investigate the effect of ferromagnetism~\cite{deshmukh,kleff}. Spin-orbit coupling and non-equilibrium excitations were studied in gold grains~\cite{bolotin1,kuemmeth,gueron}. The recent discovery of superconductivity in doped silicon at atmospheric pressure and critical temperatures of a few hundred millikelvin~\cite{busta} might further facilitate the development of mesoscopic superconducting devices. However, the competition between superconductivity and ferromagnetism has not been investigated so far.

Here we review the effects of electron-electron interactions on the ground-state spin and transport properties of ultra-small metallic grains. Our analysis is based on an effective Hamiltonian for chaotic or disordered systems that combines a BCS-like pairing term and a Stoner-like spin exchange term. This so-called universal Hamiltonian~\cite{kur,alei} is valid in the limit of a large Thouless conductance. This universal Hamiltonian and its solution are described in Sec.~\ref{model}. In Sec.~\ref{ground-state} we present the phase diagram of the ground-state spin and discuss a regime in which superconductivity and ferromagnetism coexist. In Sec.~\ref{conductance} we review the mesoscopic fluctuations of the tunneling conductance through an almost-isolated metallic grain. In particular, we discuss signatures of the coexistence of pairing and exchange correlations in the conductance peak height and peak spacing statistics. We also propose specific materials for which such mesoscopic coexistence might be observed experimentally.

\section{Model}\label{model}

We consider a chaotic metallic grain with a large dimensionless Thouless conductance. The low-energy excitations of such a grain are described by an effective universal Hamiltonian~\cite{kur,alei}
\begin{eqnarray}
\label{origH}
\hat H=\sum_{k\sigma}\hspace{-0.0cm}\epsilon_{k} c_{k\sigma}^\dagger c_{k\sigma}
-G \hat P^\dagger \hat P  -  J_s \hat{\bf S}^2 \;,
\end{eqnarray}
where $c_{k\sigma}^\dagger$ is the creation operator for an
electron in the spin-degenerate single-particle level $\epsilon_k$ with spin up
($\sigma=+$) or spin down ($\sigma=-$). The first term on the r.h.s. of (\ref{origH}) describes the single-particle Hamiltonian of an electron in the grain (i.e., kinetic energy plus confining single-particle potential). The second term is the pairing interaction with coupling constant $G$ and the pair creation operator  $P^\dagger=\sum_i c_{i+}^\dagger c_{i-}^\dagger$. The third term is an exchange interaction where  $\hat{\bf S}=\sum_{k\sigma\sigma'}c_{k\sigma}^\dagger {\bf \tau}_{\sigma\sigma'}c_{k\sigma'}$ is the total spin operator ($\tau_i$ are Pauli matrices) and $J_s$ is the exchange coupling constant. Estimated values of $J_s$ for  a variety of materials were tabulated in Ref.~\cite{gorok}. In Eq.~(\ref{origH}) we have omitted the charging energy term $e^2 \hat N^2/2C$ ($C$ is the capacitance of the grain and $\hat N$ is the number of electrons) since it is constant for a grain with a fixed number of electrons.

The universal Hamiltonian (\ref{origH}) conserves the total spin of the grain, i.e., $[\hat {H},\hat{\bf S}]=0$. Consequently, each eigenstate has a well-defined total spin $S$ and spin projection $M$. The pairing interaction scatters pairs of spin up/down electrons from doubly occupied to empty orbitals. Therefore the pairing interaction does not affect the singly occupied levels. This is known as the blocking effect and the singly occupied levels are also referred to as `blocked levels'. On the other hand, these singly occupied levels are the only levels that contribute to the exchange interaction.  Thus, each eigenstate of (\ref{origH}) factorizes into two parts. The first part $\vert {\cal U}\rangle$ is a zero spin eigenstate of the reduced BCS Hamiltonian $\sum_{k \sigma}\hspace{-0.0cm}\epsilon_k c_{k\sigma}^\dagger c_{k\sigma} -G P^\dagger P$, and is described as a superposition of Slater determinants that are constructed from the subset ${\cal U}$ of empty and doubly occupied levels. The second part of the eigenstate, $\vert {\cal B}, \gamma, S,M \rangle $, is obtained by coupling the set of singly occupied levels ${\cal B}$, each carrying spin $1/2$, to total spin $S$ and spin projection $M$~\cite{rupp,tureci}. Here, $\gamma$ denotes a set of quantum numbers distinguishing between eigenstates with the same spin and singly occupied levels. For a given set $\cal{B}$ of $b$ singly occupied levels, the allowed spin values vary between $S=0$ ($S=1/2$) for even (odd) number of electrons and $S=b/2$. Each of these spin values has a degeneracy of
\begin{eqnarray}
d_b(S) = {b \choose S+ b/2} - {b \choose S+1+ b/2}\,.
\end{eqnarray}
The complete set of eigenstates is then given by
\begin{eqnarray}
\vert i\rangle=\vert {\cal U},{\cal B}, \gamma, S,M \rangle\,.
\end{eqnarray}

The reduced pairing Hamiltonian is characterized by a coupling constant $G$. However, the physical parameter that determines the low-energy spectrum of the grain (for $J_s=0$) is $\Delta/\delta$, where $\Delta$ is the bulk pairing gap and $\delta$ the single-particle mean level spacing. We can truncate the total number of levels from $N_o$ to $N_r < N_o$,  and renormalize $G$ such that the low-energy spectrum of the grain remains approximately the same. For a picketfence spectrum, the renormalized coupling constant is given by
\begin{eqnarray}\label{renorm}
\frac{G_r}{\delta}=\frac{1}{{\rm arcsinh}\left(\frac{N_r+1/2}{\Delta/\delta}\right)}\;.
\end{eqnarray}
The exchange interaction only affects the singly occupied levels, and we expect the renormalization (\ref{renorm}) to hold as long as the number of singly occupied levels is small compared with $N_r$.  In practice, we study the relevant observable as a function of truncated bandwidth $N_r$ and make sure that its value has converged for the largest bandwidth $N_r$ we can calculate.

\section{Ground-State Phase Diagram}\label{ground-state}

In this section we study the ground-state spin of the grain as a function of $J_s/\delta$ and $\Delta/\delta$. We find the lowest energy $E(S)$ for a given spin $S$ and then minimize $E(S)$ with respect to $S$.  The ground-state spin of the grain is determined by the competition between various terms in the Hamiltonian (\ref{origH}). The one-body part and pairing interaction favor minimal spin, while exchange interaction favors a maximally polarized state.

\begin{figure}
\centerline{\includegraphics[width=3.5in]{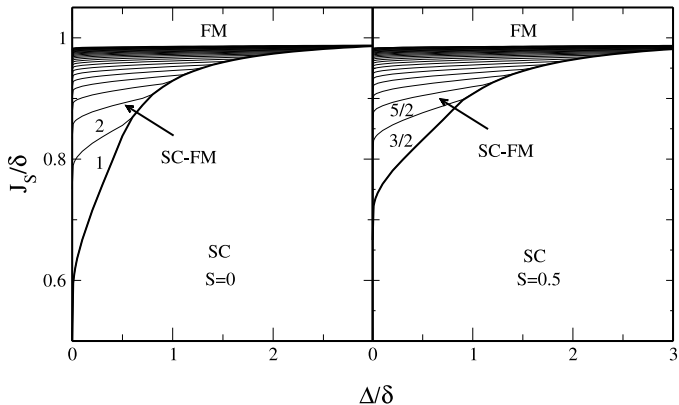}}
\caption{ Ground-state phase diagram of a grain with picketfence single-particle spectrum in the $J_s/\delta$--$\Delta/\delta$ plane for an even (left panel) and odd (right panel) number of electrons. Numbers denote the spin values in the corresponding sectors. We observe an intermediate regime (SC-FM) in which the ground state is partly polarized and partly paired. Taken from Ref.~\cite{schmidt}.}
\label{phase}
\end{figure}

The ground-state phase diagram in the $\Delta/\delta$--$J_s/\delta$ plane of a grain with a picketfence single-particle spectrum is shown in Fig.~\ref{phase}.  We find three different phases: a superconducting phase (SC) where the number of pairs is maximal and $S$ is minimal, a ferromagnetic phase (FM) where the system is fully polarized $S=N/2$, and an intermediate regime (SC-FM) where exchange and pairing correlations coexist. The ground-state wave function with spin $S$ in the coexistence regime is described by  $b=2S$ singly occupied levels closest to the Fermi energy while the remaining electrons are paired.

The coexistence regime is bounded by two critical values $J_s^{(1)}/\delta$ and $J_s^{(2)}/\delta$ of the exchange interaction that are function of $\Delta/\delta$. The lower value $J_s^{(1)}/\delta$ is a monotonically increasing function of $\Delta/\delta$ (stronger exchange is required to polarize a grain with stronger pairing correlations), while the higher value $J_s^{(2)}/\delta$ depends only weakly on $\Delta/\delta$.

It is interesting to follow the dependence of the ground-state spin as a function of the exchange coupling constant $J_s/\delta$ for a fixed value of $\Delta/\delta$. In the absence of pairing ($\Delta=0$), this dependence follows a stepwise behavior known as the mesoscopic Stoner staircase~\cite{kur}, where a transition from spin $S$ to spin $S+1$ occurs for an exchange coupling of
\begin{eqnarray}
\frac{J_s}{\delta}=\frac{2S+1}{2S + 2}\qquad\mbox{at}\quad \Delta=0 \;.
\end{eqnarray}
The first step occurs at $J_s/\delta=0.5$ (where the ground-state spin increases from $S=0$ to $S=1$) and continues up to $J_s=0.75$ (where the $S=1$ to $S=2$ transition takes place). In the presence of pairing, the first step is shifted to higher values of the exchange and the Stoner staircase is compressed. For $\Delta/\delta < 0.6$, all steps have a height of $\Delta S=1$.  However, for  $0.6 < \Delta/\delta < 0.8$, the first step has a height of  $\Delta S=2$, describing a spin jump from $S=0$ to $S=2$. This first step starts at $J_s/\delta \approx 0.87$ and ends at $J_s/\delta \approx 0.9$.  The height of the first-step spin jump increases at larger values of $\Delta/\delta$. All subsequent steps are of height one.

Spin jumps also occur when superconductivity in metallic grains breaks down in the presence of a sufficiently large external Zeeman field~\cite{BD97}. This ``softened'' first-order phase transition from a superconductor to a paramagnet was explained qualitatively using a finite-spin BCS approximation.

In the presence of exchange correlations, spin jumps are predicted to occur at $J_s/\delta > 0.87$. Such exchange coupling values are significantly larger than the exchange coupling values of most metals (see Fig.~9 in Ref.~\cite{gorok}). Moreover, the exchange strength is an intrinsic material property and is difficult to tune experimentally. In Ref.~\cite{schmidt} we have shown that the coexistence regime can be tuned to experimentally accessible values of the exchange interaction by applying an external Zeeman field.

\section{Conductance}\label{conductance}

The determination of the ground-state spin of a grain is a difficult measurement. It is then desirable to identify signatures of coexistence of superconductivity and ferromagnetism in a quantity that is directly measurable, e.g., the conductance.    Furthermore, the universal Hamiltonian (\ref{origH}) is only valid for chaotic (or disordered) grains, in which mesoscopic fluctuations are generic. Therefore, in order to make quantitative predictions for the experiment it is necessary to include the effect of mesocopic fluctuations. In this section we discuss the mesoscopic fluctuations of the tunneling conductance for an almost-isolated metallic grain in the Coulomb blockade regime. We find signatures of  the coexistence of pairing and exchange correlations in the conductance statistics. Since the tunneling conductance can be measured in a single-electron transistor that uses the metallic grain as an island, our results are directly relevant for the experiment.

We consider grains that are weakly coupled to external leads. In the regime of sequential tunneling $\delta, T \gg \Gamma$ ($\Gamma$ is a typical tunneling width). Assuming the charging energy to be much larger than temperature ($e^2/2C \gg T$), the conductance displays a series of sharp peaks as a function of gate voltage.  The $N$-th conductance peak describes a tunneling event in which the number of electrons in the dot changes from $N$ to $N+1$. and is determined by the many-body energies and transition rates between eigenstates of the $N$ and $N+1$ electrons. The conductance peak height and peak position are determined by solving a system of rate equations~\cite{al04}.

Here we present results for the conductance peak spacing and peak height statistics for an experimentally accessible temperature of $T=0.1\,\delta$~\cite{RBT97}.  In the absence of an external magnetic field, the single-particle Hamiltonian is described by the Gaussian orthogonal ensemble (GOE) of random matrices. For each random matrix realization of the one-body Hamiltonian, we use the Lanczos method to find the five lowest eigenstates of the many-body Hamiltonian (\ref{origH}). The calculations are carried out for a truncated bandwidth $N_r=8$ and electron numbers $N=16, 17, 18$ and $19$. Using the many-body energies and wave functions, we calculate the tunneling matrix elements for the corresponding three tunneling events. We then solve the system of rate equations and determine the conductance as a function of gate voltage. The peak position and height are determined numerically. To ensure good statistics, the above procedure is repeated for $4000$ realizations of the one-body Hamiltonian.

\subsection{Peak spacing statistics}

\begin{figure}
\centerline{\includegraphics[width=3in]{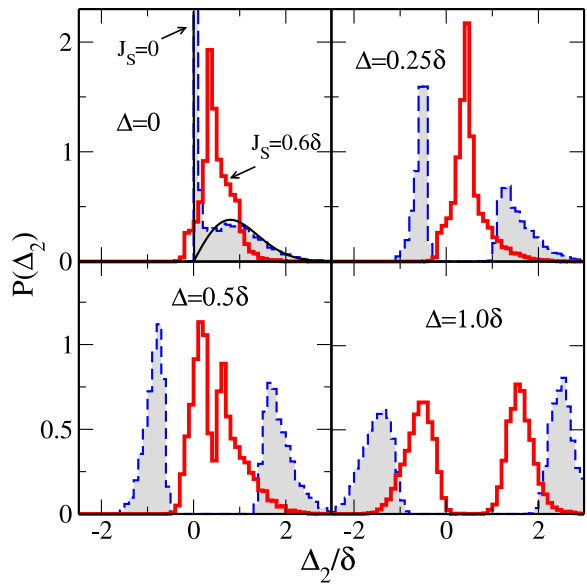}}
\caption{Peak spacing distributions at $T=0.1\,\delta$ for several values of  $\Delta/\delta$. Results are shown for both $J_s=0$ (dashed, grey-filled histograms) and $J_s=0.6\,\delta$ (solid histograms). For $\Delta=0$ we also compare with the analytic result~\cite{JS92} at $T \ll \delta$ and $J_s=0$ (solid line). The bimodality of the distribution at $\Delta=J_s=0$ is suppressed at finite exchange ($J_s=0.6 \,\delta$) but is restored for $\Delta/\delta=0.5$.  Taken from Ref.~\cite{schmidt2}.}
\label{pspa}
\end{figure}

The peak spacing distribution is shown in Fig.~\ref{pspa}, where the spacing is measured relative the constant charging energy. For $\Delta=J_s=0$ and at low temperatures, this distribution is bimodal because of the spin degeneracy of the single-particle levels~\cite{al00}. The exchange interaction induces mesoscopic spin fluctuations and suppresses this bimodality (see top left panel of Fig.~\ref{pspa}). This is known from the case of semiconductor quantum dots.

Pairing correlations can restore bimodality. For a moderate exchange value of $J_s=0.3\,\delta$, bimodality is restored for a relatively weak pairing strength of $\Delta = 0.25\,\delta$. For $J_s=0.6 \,\delta$, this bimodality is suppressed but reappears at $\Delta/\delta=0.5$ (see bottom left panel of Fig.~\ref{pspa}). The left part of the peak spacing distribution describes even-odd-even (E-O-E) tunneling events (the parity refers to the number of electrons), and its right part  describes odd-even-odd (O-E-O) transitions.

The bimodality of the peak spacing distribution in the presence of strong pairing correlations can be understood qualitatively in the $T=0$ fixed-spin BCS approximation~\cite{schmidt}.  For an E-O-E transition, the first conductance peak corresponds to the blocking of an additional single-particle level, while the second conductance peak corresponds to the removal of this blocked level by creating an additional Cooper pair. This leads to the estimate $\Delta_2^{\rm EOE}\approx -2\Delta+\frac{3}{2}J$. In a O-E-O tunneling sequence, these two events are reversed and we find $\Delta_2^{\rm OEO}\approx 2\Delta-\frac{3}{2}J$.
The contribution of the exchange interaction in these estimates is straightforward  because, in the limit of strong pairing, the ground-state spin is always $S=0$ ($S=1/2$) for an even (odd) number electrons. The difference of these two peak spacing values is
\begin{eqnarray}
\delta\Delta_2=\Delta_2^{\rm OEO}-\Delta_2^{\rm EOE}\approx 4\Delta-3J\,,\quad\Delta\gg\delta \;,
\end{eqnarray}
and bimodality becomes more pronounced when $\Delta/\delta$ increases. Since the exchange interaction strength for most metals is smaller or comparable to $J_s\sim 0.6 \, \delta$, exchange correlations are insufficient to suppress the bimodality in the presence of strong pairing correlations.

\begin{figure}
\centerline{\includegraphics[width=3in]{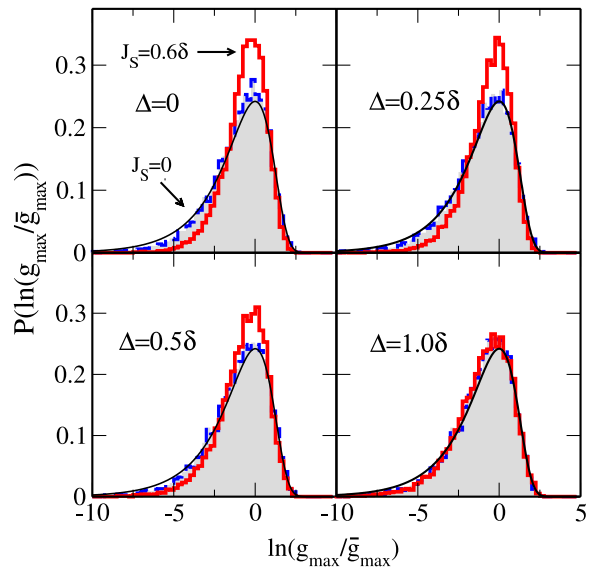}}
\caption{Peak height distributions at $T=0.1\,\delta$ for several values of $\Delta/\delta$. We show results for $J_s=0$ (dashed, grey-filled histograms) and $J_s=0.6\,\delta$ (solid histograms). The solid line is the analytic distribution at $T\ll\delta$ and $J_s=0$~\cite{JS92}. Since the conductance peak height $g_{\rm max}$ fluctuates over several order of magnitude, we show the distributions as a function of $\ln(g_{\rm max}/\bar g_{\rm max})$ where $\bar g_{\rm max}$ is the average conductance peak height.  Taken from Ref.~\cite{schmidt2}.}
\label{pheight}
\end{figure}

\subsection{Peak height statistics}

The peak height distribution is shown in Fig.~\ref{pheight}. For $\Delta=J_s=0$, this distribution is known analytically at $T \ll \delta$ ~\cite{JS92}, and, in the absence of an external magnetic field (GOE statistics), diverges at small values of the conductance. Finite temperature and exchange interaction have a similar effect on the peak height distribution; they both reduce the occurrence of small conductance values. While thermal fluctuations open an energy window in which states become available for tunneling and thus can contribute to the conductance, the exchange interaction increases the many-body density of states around the Fermi energy and makes otherwise high-lying non-zero spin states available for tunneling. This effect is shown in Fig.~\ref{pheight}. For $\Delta=0$ we clearly observe that a finite exchange interaction suppresses the peak height distribution at small conductance values. The same behavior was observed for the Gaussian unitary ensemble (GUE) ensemble in semiconductor quantum dots, where the pairing interaction can be ignored and a closed solution for the conductance is available~\cite{rupp}. There, the suppression of probability at small conductance values by the inclusion of exchange interaction leads to better agreement with the experimental results. Another signature of exchange correlations is the suppression of peak height fluctuations as described by $\sigma(g_{\rm max})/\bar g_{\rm max}$ (the ratio between is the standard deviation and average value of the peak height)~\cite{schmidt2}.

However, it is important to note that at very low temperatures, e.g., $T=0.01\,\delta$, the small conductance values are no longer suppressed by an exchange interaction. For such very low temperatures, the increase in the density of states near the Fermi energy by the exchange interaction does not affect the conductance since only the ground state contributes significantly to transport.

The pairing interaction leads to an excitation gap that pushes states with large spin to higher energies. Thus, already for  $\Delta=0.5\,\delta$ we observe less suppression of the small conductance values by an exchange of $J=0.6\,\delta$. When the pairing interaction is strong enough to suppress all spin polarization of low-lying states (e.g., $\Delta \geq 1.0\,\delta$ for $J_s <0.6\,\delta$), the peak height distribution becomes essentially independent of the exchange interaction and diverges again for small conductance values.

For $\Delta/\delta=0.5$ and $J_s/\delta =0.6$, we observe a signature of pairing correlations in the peak spacing distribution (bimodality) and a signature of exchange correlations in the peak height distribution (suppression of the probability of small conductance values). We can interpret these results to describe the mesoscopic coexistence of pairing and exchange correlations. A candidate for this mesoscopic coexistence is vanadium, which has $J_s/\delta \sim 0.57-0.63$~\cite{gorok} and is superconducting in the bulk. Another candidate is platinum, which has $J_s/\delta \sim 0.59-0.72$ and is superconducting in granular form.

\section{Conclusion}

We reviewed the competition between superconductivity and ferromagnetism in ultra-small metallic grain. In particular, we presented the ground-state phase diagram in the $J_s/\delta-\Delta/\delta$ plane, and discussed a coexistence regime of superconductivity and ferromagnetism. This regime is characterized by spin jumps that are greater than unity. We also discussed the transport properties of the grain in its Coulomb blockade regime of weak coupling to leads and described the statistics of the conductance peaks in the presence of both pairing and exchange correlations.  Of particular interest is a regime in which pairing and ferromagnetic correlations coexist. Such a regime is defined by the simultaneous occurrence of bimodality in the peak spacing distribution (caused by pairing correlations) and the suppression of the peak height distribution at small conductance values (caused by ferromagnetic correlations). 

The exchange interaction strength is a material constant and might be difficult to tune experimentally. Alternatively, the coexistence regime can be controlled by an external Zeeman field while measurements are carried out at a fixed exchange interaction strength.

\acknowledgements

This work was supported in part by the U.S. DOE grant No. DE-FG-0291-ER-40608. Computational cycles were provided by the NERSC high performance computing facility at LBL and the Yale Bulldog cluster.

\end{document}